\documentclass{aa} 
\usepackage{epsf}
\usepackage{graphicx}
\usepackage{latexsym}

\newcommand{\mr}{$M_{\rm R}$}

\newcommand{\Fnu}{$F_{\nu}$}

\newcommand{\vi}{$V-I$}
\newcommand{\ebv}{$E_{\rm B-V}$}

\begin{document}
\thesaurus{05(02.01.1, 11.07.1, 11.19.3, 12.03.2, 13.07.1)}
\title{VLT observations of GRB\,990510 and its environment
\thanks{Based on observations collected at the European Southern
Observatory, Chile, programs 63.P-0003, 63.H-0009, and 63.O-0567 (B).}
}
\author{K. Beuermann
\inst{1}
\and F.V. Hessman
\inst{1}
\and K. Reinsch
\inst{1}
\and H. Nicklas
\inst{1}
\and P.M. Vreeswijk
\inst{2}
\and T.J. Galama
\inst{2}
\and E. Rol
\inst{2}
\and J. van Paradijs
\inst{2,3}
\and C. Kouveliotou
\inst{4}
\and F. Frontera 
\inst{5,6}
\and N. Masetti
\inst{5}
\and E. Palazzi
\inst{5}
\and E. Pian
\inst{5}
}
\offprints{beuermann@uni-sw.gwdg.de}
\institute{Universit\"ats-Sternwarte, Geismarlandstr. 11, D-37083
G\"ottingen, Germany 
\and Astronomical Institute `Anton Pannekoek',University of Amsterdam,
\& Center for High Energy Astrophysics, Kruislaan 403, 1098 SJ
Amsterdam, The Netherlands
\and Physics Department, University of Alabama in Huntsville, Huntsville AL 35899, USA NASA Huntsville/Alabama
\and Universities Space Research Association and  NASA/MSFC, Code ES-84, Huntsville AL 35812, USA
\and Istituto Tecnologie e Studi Radiazioni Extraterrestri,
CNR, via Gobetti 101, I--40129 Bologna, Italy
\and Dipartimento di Fisica Universit\'a di Ferrara, Via Paradiso 12, 44100 Ferrara, Italy }
\date{Received August 12, 1999 / Accepted September 1, 1999}
\authorrunning{K. Beuermann et al.}
\titlerunning{VLT observations of GRB\,990510}
\maketitle
\begin{abstract}
We present BVRI photometry and spectrophotometry of GRB\,990510
obtained with the ESO VLT/{\it Antu} telescope during the late decline
phase.  Between days 8 and 29 after the burst, the afterglow faded
from $R$ = 24.2 to $\sim26.4$. The spectral flux distribution and the
light curve support the interpretation of the afterglow as synchrotron
emission from a jet. The light curve is consistent with the optical
transient alone but an underlying SN with maximum brightness $R >
27.4$ or a galaxy with $R > 27.6$ (3-$\sigma$ upper limits) cannot be
ruled out.  To a 5-$\sigma$ detection threshold of $R=26.1$, no galaxy
is found within 6'' of the transient. A very blue $V \simeq 24.5$
extended object which may qualify as a starburst galaxy is located
12'' SE, but at unknown redshift.

\end{abstract} 

\section{Introduction} 

Gamma-ray bursts (GRBs) have their origin in highly dynamical
processes which result in relativistic blast waves (e.g. Piran
1999). The currently best candidates are mergers of compact stars or
the collapse of massive stars. The blast wave and its interaction with
the surrounding interstellar or circumstellar matter creates an
optical afterglow which has been observed for eleven GRBs thus
far. For seven of these, the host galaxies have been found and for
five the redshift could be measured (Hogg \& Fruchter 1999). While the
final cataclysms of massive stars occur within or near the host
galaxies, less massive systems like binary neutron stars can be
expelled from their hosts at large velocities and suffer the merging
event at distances from the hosts exceeding a Mpc (Fryer et al. 1999).

GRB\,990510 was detected by BATSE, Ulysses, and BeppoSAX as a very
bright burst lasting 68\,s (see Wijers et al. 1999 for the burst
profile and a first summary). The optical afterglow was discovered by
Vreeswijk et al. (1999a) and subsequently followed by numerous
observers. An early spectrum located the optical transient (OT) at a
redshift $z \ge 1.62$ (Vreeswijk et al. 1999b). GRB\,990510 is the
first burst to show a clearly defined achromatic break in the optical
light curve (Harrison et al. 1999, Israel et al. 1999, Stanek et
al. 1999) which was readily interpreted as firm evidence for beaming
and a total energy release substantially less than the isotropic value
of $3\times10^{53}$\,ergs.  In this Letter, we report observations
with the ESO VLT of the late decline of the optical afterglow at two
epochs.

\section{Observations}

GRB\,990510 was observed with ESO's Very Large Telescope (VLT) Unit\,1
{\it Antu} equipped with the Focal Reducer Low Dispersion Spectrograph
(FORS1) between May 14 and 18, 1999, and again between June 8 and 11,
1999 (3.8--8.0 and 28.7--31.6 heliocentric days after burst, hereafter
HDAB).  We performed photometry in the Bessel B,V,R, and the Gunn I
bands, supplemented by long-slit spectrophotometry. The scale was
0.2''/pixel in all observations except for the B, R images on June 8,
1999 which were taken in the high-resolution mode of FORS1 with
0.1''/pixel. The data were reduced with standard {\it ESO-MIDAS}
procedures. Photometry was derived relative to the comparison stars
given by Pietrzynski \& Udalski (1999), with stars A and B near the
optical transient (Fig. 3) serving as secondary standards (Table
1). Due to wind buffeting, the point spread function (PSF) was
slightly non-circular in some images, so magnitudes were determined by
a special PSF-template routine. Table 1 provides a summary of the
observations with times given in heliocentric Julian days (HJD) and
HDAB. 1-$\sigma$ internal statistical errors are quoted for the
photometric magnitudes.

Long-slit spectrophotometry was performed on May 14 and 16, 1999.  The
slit was oriented at a position angle $-15^\circ$ to include objects C
south and A north of the OT (Fig. 3) and, coincidentally, also two M
stars 12'' north of the OT ($R = 22.1,23.5$, outside of Fig. 3). The late
spectroscopy of the OT is reported elsewhere (Vreeswijk et al. 1999).

\begin{table}[b]
\caption[ ]{Bessel photometry of stars A, B, C, and D in Fig. 3. The
errors quoted refer to the internal statistical uncertainties
only. The primary standard is star No. 3 of Pietrzynski \& Udalski
(1999) with $B = 17.88, ~V = 17.01, ~R = 16.50$, and~~$I = 16.09$.}
\begin{tabular}{cc@{\hspace{3mm}}c@{\hspace{3mm}}c@{\hspace{3mm}}c}
\noalign{\smallskip} \hline \noalign{\smallskip}
Star & B & V & R & I \\ 
\noalign{\smallskip} \hline \noalign{\smallskip}
A & $21.38\pm0.01$ & $20.03\pm0.01$ & $19.17\pm0.01$ & $18.45\pm0.02$ \\
B & $21.23\pm0.01$ & $20.03\pm0.01$ & $19.27\pm0.01$ & $18.68\pm0.02$ \\
C & $23.70\pm0.02$ & $22.63\pm0.02$ & $22.02\pm0.02$ & $21.48\pm0.02$ \\
D & $26.60\pm0.95$ & $24.57\pm0.09$ & $23.64\pm0.06$ & $22.69\pm0.04$ \\
\noalign{\smallskip}\hline
\end{tabular}
\end{table}

The seeing encountered during the observations was never better than
0.8'' (Table 2). This resulted in 5-$\sigma$ detection thresholds of $B$ =
26.6 on June 8, $V$ = 25.6 on May 18, $R$ = 26.1 on June 8, and $I$ =
25.0 \mbox{on May 14 and June 11}.

\section{Results}

\subsection{Spectrophotometry}

Our slit spectra of objects A and C reveal their stellar nature.
Object A is a dK$6\pm1$ star of the disk population. Its brightness
(Table 1) places it at a distance of $d \simeq 2$\,kpc, outside the
galactic dust layer. The spectral type implies an intrinsic
\vi{}$=1.40\pm0.13$ and \ebv{}$=0.14\pm0.09$, consistent with
\ebv{}$=0.20\pm0.03$ quoted by Stanek et al. (1999). Object C,
initially suspected to be the host galaxy, is pointlike. Its spectrum
is included in Fig. 1. The probable presence of Mg\,I\,$\lambda 5167$
suggests that it is an early dK star at a distance of $\sim 15$\,kpc.
The photometry of objects B and D suggests they are stars of spectral
types dK5 and dM0, respectively (Table 1).

Figure 1 also shows the extinction-corrected spectrophotometric flux
distribution of the OT on HDAB 3.9 when the OT had $R$ = 21.98,
corrected with \ebv{} = 0.20 (histogram; binned in intervals of
200\AA). The solid dots denote the photometric result. For the
interval of 4900--9000\AA{}, the flux distribution can be described by
a power law \Fnu \,$\propto\,\nu^{-\beta}\propto\,\lambda^{\beta}$
with $\beta = 0.55\pm0.10$. For higher frequencies, the spectral flux
distribution steepens, as noted already by Stanek et al. (1999). The
steepening may be intrinsic to the source or indicate additional
absorption outside our Galaxy.

\begin{figure}[t]
\includegraphics[width=8.8cm]{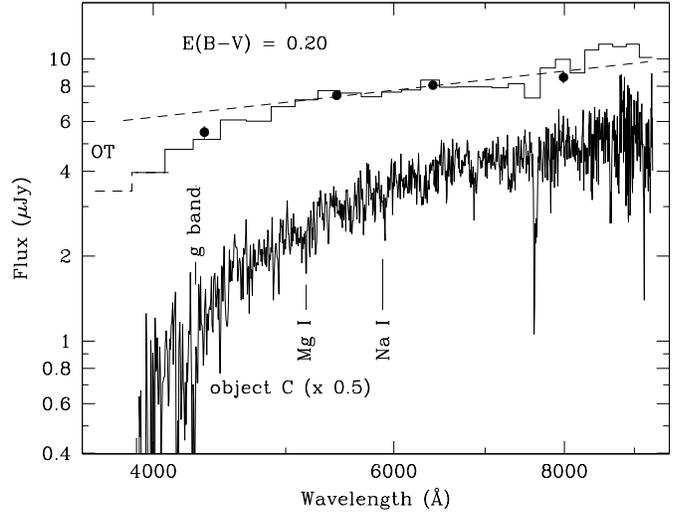}
\caption[]{Spectral flux distribution of the OT of GRB\,990510,
de-reddened with \ebv{} = 0.20. The solid points refer to the mean
magnitudes of the OT on HDAB 3.9, the histogram presents our
spectrophotometry averaged over 200\AA\ bins. The dashed line is the
power-law fit to $\lambda > 4900$\AA\ with slope $\beta = 0.55$ for a
power law \Fnu \,$\propto\,\nu^{-\beta}\propto\,\lambda^{\beta}$. 
Shown below is the spectrum of the $V = 22.6$ early K-star C, 2''
south of the OT, shifted downward by a factor of two.}
\vspace{-1mm}
\end{figure}

\vspace{-1mm}
\subsection{Light curve}

Our BVRI photometry provides colour information late in the afterglow
which is complementary to the ample information gathered earlier by
other observers. Around HDAB 1.5, the light curve was found to change
from an early power law $F = k\,t^{-\alpha}$ with $\alpha = 0.8$ to a
steeper power law with index 2.2 (Harrison et al. 1999, Israel et
al. 1999, Stanek et al. 1999, and references therein).

In the HDAB 0.6--1.1 interval, i.e. before the break, the OT had
colours $B-V = 0.57\pm0.02, ~V-R = 0.41\pm0.01$, and $R-I =
0.47\pm0.01$. The early spectral flux distribution was possibly
slightly bluer, as judged from the Mt. Stromlo data (Harrison et
al. 1999) which yield $V-R = 0.31\pm0.03$ at HDAB 0.15 and $V-R =
0.36\pm0.05$ for HDAB 0.4. Throughout the further evolution, however,
there is no evidence for a significant change in colours: (i) in the
HDAB 1.6--2.0 interval are $B-V = 0.61\pm0.07, ~V-R = 0.42\pm0.03$,
and $R-I = 0.40\pm0.02$, consistent with the colours before the break;
(ii) on HDAB 3.85, well after the brak, our BRI photometry yields $B-R
= 0.98\pm0.07$ and $R-I = 0.49\pm0.06$; and (iii) the latest colour
information available, for HDAB 5.7--8.0, yields a mean $V-R =
0.37\pm0.03$ (all colours derived from magnitudes corrected for
long-term trends). These results demonstrate that the decay of the
afterglow is achromatic in VRI at least for the interval HDAB 0.6--8.0
and in BVRI for HDAB 0.8--3.9. We feel justified, therefore, in
converting all measurements to equivalent $R$-magnitudes and fitting a
single grand total light curve (Fig.~2. upper panel). The final mean
colours used in this conversion were determined iteratively and are
given below.

The errors attached to the individual data points are the statistical
errors derived from the noise in the CCD images. Systematic zero point
errors are typically quoted as 0.02--0.03\,mag, consistent with the
internal scatter of $\sim0.02$\,mag and the lack of obvious time
variability among the 101 (of a total of 163) data points in the HDAB
0.6--1.1 interval.  While short term variability is certainly small, a
systematic search for such variability among {\it all} photometric
data is still pending.  In order to see if systematic zero point
errors affect the fit, we also \mbox{analysed} a data set for which
0.03 mag was quadratically added to the statistical errors before
fitting the light curve.

\begin{figure*}[t]
\begin{center}
\includegraphics[width=14.5cm]{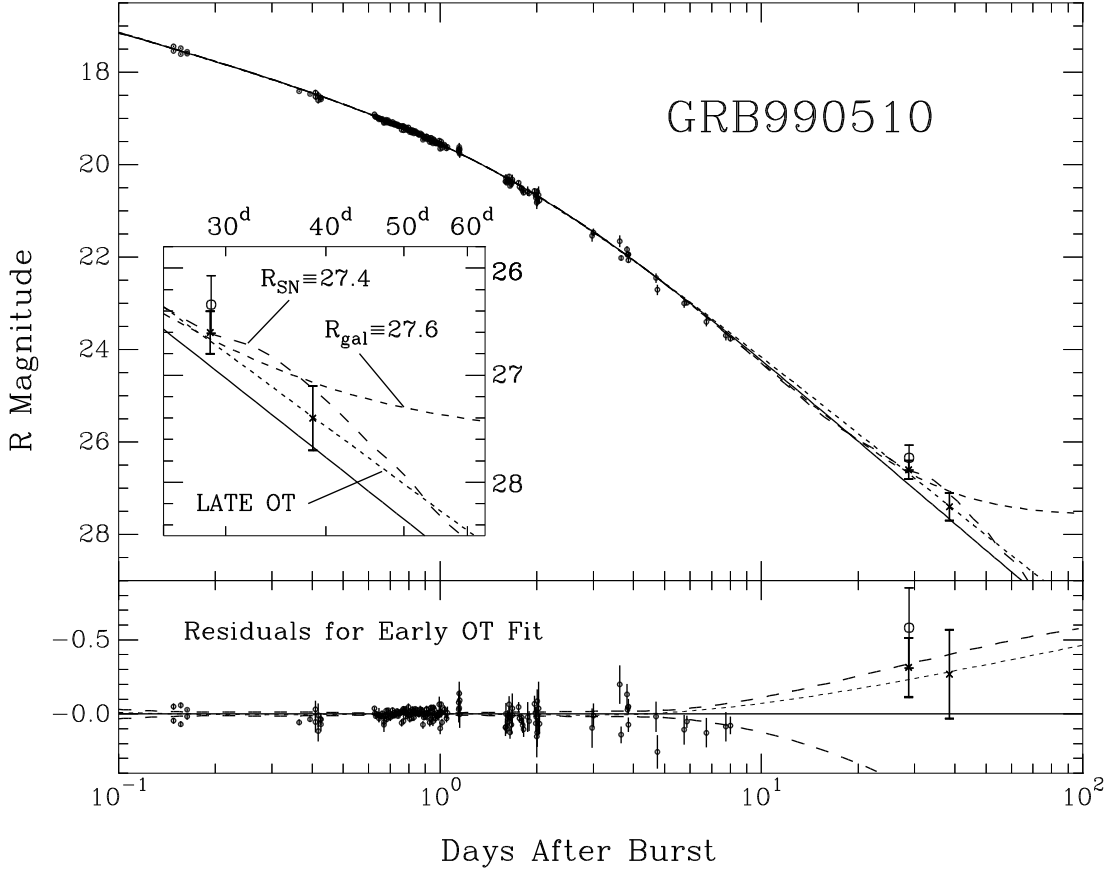}
\caption[]{{\it Top: } Light curve of the OT of GBR990510, with all
published measurements converted to the $R$-band. The solid curve
indicates the fit to HDAB$ < 8.0$ data. {\it Insert:} Fits to all
available data assuming the OT alone (Late OT, dotted), the OT with a
3-$\sigma$ upper limit flux to a supernova (long dashes) or a host
galaxy (short dashes), $\times$ HST, $\circ$ late VLT data. {\it Bottom:
} residuals relative to a pure OT light curve fit to the HDAB $< 8$
data points. The 3-$\sigma$ bounds (dashed lines) show the
uncertainties in the extrapolation to late times. The dotted line
indicates the pure OT fit including the three late data points. }
\end{center}
\end{figure*}

Burst models (Piran 1999, Rhoads 1999, Sari et al. 1999) explain the
afterglow as synchrotron emission of shock-accelerated electrons
injected into an expanding me\-dium with a power law spectrum
$E^{-p}$. In jet models, the time dependent spectral flux at
frequencies below the cooling break varies as $F_{\nu}(t) \propto
\nu^{-\beta}\,t^{-\alpha}$, with $\beta = (p-1)/2$ independent of time
and $\alpha = 3(p-1)/4$ or $\alpha = p$, depending on, respectively,
whether the opening angle of the relativistically beamed radiation
$\vartheta \simeq 1/\gamma < \theta$ early in the expansion, or
$\vartheta > \theta$ at later time when the jet has been slowed down
($\theta$ = opening angle of the jet, $\gamma$ = bulk Lorentz
factor). We choose a function

\begin{equation}
F(t) = (F_1^{-n} + F_2^{-n})^{-1/n} \qquad {\rm with} 
~ F_i \equiv k_{\rm i} t^{-\alpha_i}, ~n > 0\hfill
\end{equation}

\noindent to describe the transition between the early and late power
laws $F_1$ and $F_2$, where $F_1 = F_2$ at the transition time
\mbox{$t = t_*$}.  Eq.~(1) was also employed by Rhoads (1999) to
parameterise his numerical models and is a more general form of the
expressions used by Israel et al. (1999), Stanek et al. (1999), and by
Harrison et al. (1999), who assumed $n = 1$ and $n \simeq 1.5$,
respectively. Of the five free parameters in Eq. (1) ($k_{1,2},
\alpha_{1,2}, n$), the exponent $n$ provides a measure of the relative
width and the smoothness of the transition from $F_1$ to $F_2$.  The
extrapolation of the OT brightness from HDAB$\le 8.0$ to late times
will also depend on the choice of $n$.

\begin{table*}[t]
\caption[ ]{Log of our observations of GRB\,990510 with the ESO VLT
Antu (UT1) and the Focal Reducer Spectrograph FORS1.  HDAB is the time
of the centre of the exposure in heliocentric days after the
burst which occurred on HJD 2451308.870.  1-$\sigma$ statistical
errors are quoted. }
\begin{flushleft}
\begin{tabular}{lcrlccrccc}
\noalign{\smallskip} \hline \noalign{\smallskip}
Date of & HJD      & HDAB & Observ. & Seeing & Filter or  & Expos. & 
Flux rel. to star A & Magnitude & Equiv. R-mag\\
1999 & --2451300   &(days) &             & (arcsec) & Grating &  (sec)   & 
(\%)                  &           &                   \\ 
\noalign{\smallskip} \hline \noalign{\smallskip}
May 14 & 12.707& 3.837&Phot. &0.8&  R        & 300&$7.67\pm0.11$&$21.95\pm0.02$&\\
May 14 & 12.716& 3.846&Phot. &0.9&  I        &1200&$6.24\pm0.07$&$21.45\pm0.01$&$21.93\pm0.01$\\  
May 14 & 12.730& 3.860&Phot. &1.1&  B        & 100&$23.01\pm0.96$&$22.93\pm0.05$&$21.95\pm0.05$\\
May 14 & 12.763& 3.893&Spectr.&0.8&G300V      &1800&&&\\
       & 12.789& 3.919&Spectr.&0.8&G300I/OG590&1800&&&\\
May 16 & 14.731& 5.861&Phot. &1.2&  V        & 600&$4.41\pm0.14$&$23.39\pm0.04$&$22.99\pm0.04$\\
May 16 & 14.746& 5.876&Spectr.&1.5&G150I      &1800&&&\\
May 18 & 16.867& 7.997&Phot. &1.0&  V        &1200&$2.24\pm0.03$&$24.16\pm0.06$&$23.75\pm0.06$\\
June 8 & 37.628&28.758&Phot. &0.8&  B        &1200&$0.57\pm0.20$&$>26.7$&$>25.7$\\
June 8 & 37.656&28.786&Phot. &0.9&  R        &2400&$0.14\pm0.04$&$26.34\pm0.27$&\\
June 11 &40.485&31.615&Phot. &0.9&  I        &3000&$0.13\pm0.09$&$>25.1$&$>25.6$\\
\noalign{\medskip}\hline
\end{tabular}
\end{flushleft}
\vspace{-2mm}
\end{table*}
%
\begin{figure*}[t]
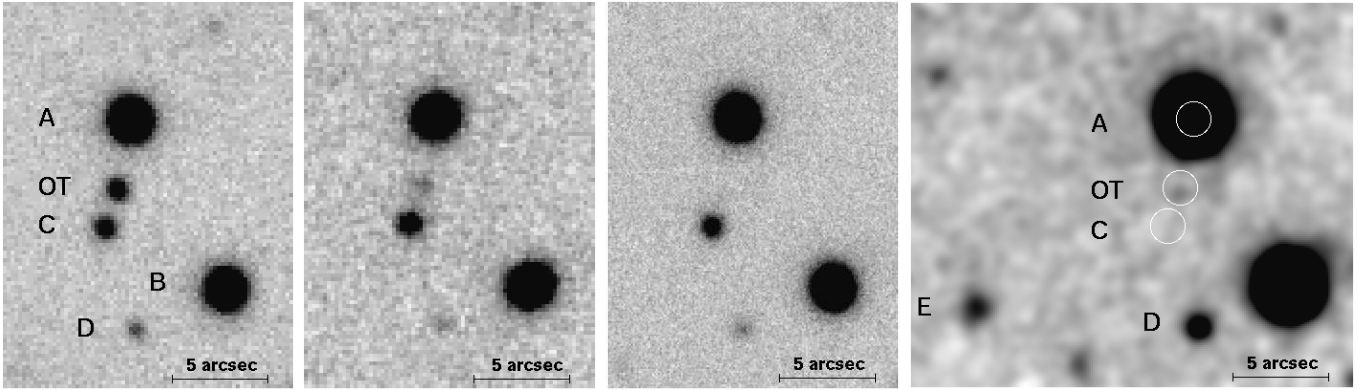

\includegraphics[width=3.9cm]{vfi06.f3a}
\includegraphics[width=3.9cm]{vfi06.f3b}
\includegraphics[width=3.92cm]{vfi06.f3c}
\raisebox{-0.3mm}{\includegraphics[width=5.95cm]{vfi06.f3d}}
\caption[]{{\it (a) -- (c), from left:} Deep $I, V$, and $R$ images of
the field of GRB\,990510 taken on May 14, May 18, and June 8 when the
OT was at $I = 21.46, V = 24.16$, and $R \simeq 26.7$. {\it (d)}
Smoothed $R$-image of June 8 with star C subtracted to better show
possible nearby faint sources. The circles at the positions of A, C
and the OT have radii of 1\,FWHM of the PSF. Object E is a 24 mag possible
starburst galaxy, 12'' SE of the OT.}
\end{figure*}

We first fit the light curve of the OT using all of the available data
up to HDAB 8.0. For the data sets with zero (0.03\,mag) systematic
errors added, we find $\alpha_1 = 0.79\pm0.04 ~ (0.81\pm0.05), ~
\alpha_2 = 2.41\pm0.16 ~ (2.31\pm0.16), ~ n = 0.87\pm0.22 ~
(1.09\pm0.34)$, and $t_* = 1.75\pm0.23 ~ (1.57\pm0.19)$ days.  The
quoted uncertainties are 1-$\sigma$ errors. The fitted $\chi^2/\nu
\equiv 333/165\,d.o.f=2.02$ shows that the fit to the unaltered data
is not perfect, but is acceptable if systematic errors of 0.03\,mag
are considered ($\chi^2/\nu = 0.83$). The observed \mbox{$n
\simeq 0.9-1.1$} differs significantly from the value $n = 0.4$ found
by Rhoads (1999) when fitting his numerical results, which predict a
rather smooth transition. The more abrupt break may be due to the
superposition of several effects, as the slowing of the jet and its
sideways spread (Harrison et al. 1999).  By fitting the shape of the
mean light curve to the fluxes in each photometric band we obtain the
magnitude normalisations corresponding to the quantities $k_i$ in
Eq.~1: $B_1 = 20.550\pm0.012$, $V_1 = 19.961\pm0.003$, $R_1 =
19.565\pm0.003$, and $I_1 = 19.075\pm0.003$. The corresponding colours are
consistent with those for HDAB 0.6--1.1. The lack of
substantial colour variations can be seen from the small scatter in
the residuals of the best fit (Fig.~2).

For comparison, for $n$ fixed at 1.0 we obtain $\alpha_1 =
0.80\pm0.02$, $\alpha_2 = 2.33\pm0.04$, $t_* = 1.63\pm 0.08$ days and
$\chi^2/\nu = 2.0$ for a fit to the unaltered data points. Since the
errors of the $\alpha$'s and of $n$ are correlated, fixing $n$ reduces
the errors in the other parameters artificially. 
Our fit to the light curve with $n \ne 1$ allows us to predict the
brightness of the OT at later times with some confidence that
uncertainties in the shape of the light curve have been accounted for
as far as possible.  We predict an OT magnitude of $R =
26.92^{-0.33}_{+0.46}$ on June 8 (HDAB 28.8), when our BR photometry
was performed.

Fruchter et al. (1999) have reported detections with the HST-STIS open
CCD on June 8.1 and June 17.9 (HDAB 28.7 and 38.5). Assuming an \Fnu
$\propto \nu^{-0.6}$ spectrum, they obtained $V = 27.0\pm0.2$ and $V =
27.8\pm0.3$, corresponding to $R = 26.6\pm0.2$ and $R = 27.4\pm0.3$,
respectively. Given the red sensitivity of the STIS-CCD, these $R$
magnitudes are quite robust. We conclude that our June 8 result and
the first of the HST detections are consistent with each other.

Assuming that all photometry including the late data points is that of
the decaying OT, we obtain nearly identical light curve parameters:
$R_1 = 19.539\pm 0.063\, (19.555\pm 0.082),\, \alpha_1 = 0.82\pm0.03
\, (0.84\pm0.03); ~\alpha_2 = 2.23\pm0.07 ~(2.18\pm0.07)$; $ n =
1.17\pm0.17 \,(1.43\pm0.26);~t_*=1.51\pm0.09 ~ (1.41\pm0.08); ~
\mbox{and} ~\chi^2/\nu = 1.87 ~ (0.81)$.  Thus, all of the available
photometry is consistent with the light curve of a pure OT.

The observed spectral index $\beta \simeq 0.55$ and the values of
$\alpha_1$ and $\alpha_2$ are in excellent agreement with the
theoretical predictions if the cooling break stays at wavelengths
shorter than the visual band until HDAB 8.0. The slope of the
electron injection spectrum then is $p = 2.1\pm0.1$ in agreement with
the conclusions drawn by Harrison et al. (1999) and Stanek et
al. (1999).

Figures 3a-c depict the late decay of the OT from an equivalent 
$R$-magnitude of 21.95 (HDAB 3.8) over 23.75 (HDAB 8.0) to ~26.4 (HDAB
28.8) (Table 2). Down to the 5-$\sigma$ thresholds for a detection at
an arbitrary position, $B \simeq 26.6, V \simeq 25.6, R \simeq 26.1$
and $I \simeq 25.0$ our images show no other object within 6'' of the
OT, besides the stars A, B, C, and D. The faint emission on June 8 at
the OT position is more easily seen when the high-resolution image
(0.1'' pixels) is slightly smoothed. Figure 3d is an enhanced version
of 3c, smoothed with a Gaussian of 0.3'' standard deviation and with
star C PSF-subtracted.  While the emission appears to be slightly
offset from the OT position by 0.4'' and may signify an underlying
object, the shift is within the statistical uncertainty expected for
such a faint source.

\subsection{The quest for the host galaxy of GRB\,990510}

Although the light curve analysis does not require another light
source in addition to the OT, such contribution can not be
excluded. One possibility is that an underlying SN-event contributes
to the June 1999 light level. Taking the $R$-band light curve of
SN1998bw (Galama et al. 1998, Iwamoto et al. 1998) adjusted to $z =
1.62$ (Bloom et al. 1999) one expects a peak magnitude of $R = 26.8$
(reddened 27.3), just consistent with the \mbox{3-$\sigma$} upper
limit of $R = 27.4$ of the fit [fitted flux relative to star A with R
= 19.17 is $(1.0\pm1.4)\times 10^{-4}$]. Forcing the fit to use a peak
value of 27.4 results in an increase in $\alpha_2$ from 2.37 to 2.71
and a correlated decrease in $n$ from 0.94 to 0.60. Alternatively, the
best-fit contribution by an underlying galaxy, which adds a constant
to the late-time light curve, yields a \mbox{3-$\sigma$} upper limit
of 27.6 (27.1 de-reddened)~[flux relative to \mbox{star A}
$(2.0\pm1.7)\times 10^{-4}$]. The forced fit with the 3-$\sigma$ upper
limit requires $\alpha_2$ and $n$ changes nearly identical to those of
the corresponding SN fit. We note that many burst models do not
require a SN event and that a host fainter than 27 mag presents no
problem with the galaxy population responsible for GRBs (Hogg \&
Fruchter 1999). Fig.~2 shows the fits obtained for the OT alone (early
data or all data) and the forced fits with the OT and the 3-$\sigma$
upper-limit SN and host contributions added: there is presently no way
to distinguish between the three models, but the fact that the early
and late OT light curve fits are so similar strongly suggests that we
have only seen the OT.

For $z = 1.62$ (Galama et al. 1999) and a standard cosmology with $H_o
= 70$\,km\,s$^{-1}$Mpc$^{-1}$ and $\Omega_o = 0.3$, the distance
modulus is 45.2. The de-reddened host magnitude $R > 27.1$ then
translates to a restframe \mr\,$> -18.4$ for a starburst spectrum
similar to NGC\,4449 (Bruzual \& Charlot 1993) and to \mr\,$> -17$
for a pure starburst spectrum. Hence, the host is intrinsically faint
if at z=1.62. While one might be willing to place the burst at a
larger $z$, there is a probable limit from the expected
Ly$\alpha$-forest or Lyman-continuum absorption. The lack of a
pronounced depression at wavelengths longer than 3600\AA{} (Fig.~1)
limits the acceptable redshift to $z \la 2.0$ for Ly$\alpha$-forest
and to $z \la 2.9$ for Lyman-continuum absorption.

The only brighter object possibly related to the OT is the one denoted
E in Fig. 3d, 12'' to the SE. With a diameter of $\sim 1''$ and
its decidedly blue colour it may qualify as a starburst galaxy. Its
de-reddened magnitudes and colours are well within the range found for
the hosts of other GRBs and OTs, $V \simeq 23.8$, $B-V \simeq 0.1, V-R
\simeq 0.3$, $R-I \simeq 0.4$, assuming \ebv{} = 0.20. The observed
spectral flux distribution is typical of the UV spectrum of a
starburst or irregular galaxy (e.g. Bruzual \& Charlot 1993). If this
is actually a galaxy at $z = 1.62$, it would be separated by at least
100 kpc from the OT. This would not be a problem if the GRB progenitor
were a low-mass system ejected from this galaxy (Fryer et
al. 1999). The absence of an associated SN or host galaxy at the OT
position would support such a scenario, while the suspected connection
of long gamma-ray bursts with massive progenitors does not.

\section{Conclusions}

The recent observations of the optical transient of \linebreak
\mbox{GRB\,990510} strongly support the synchrotron model for
gamma-ray bursts. The pre-break light curve and the $\lambda >
4900$\,\AA{} optical flux distribution are consistent with the
adiabatic cooling of relativistic electrons with an $E^{-p}$ injection
spectrum with $p \simeq 2.1$. The steep late decline is a firm
indicator of the presence of a slowing jet (Meszaros \& Rees 1999,
Rhoads 1999, Sari, et al. 1999).

The light level on June 8 is quite consistent with being solely due to
the transient, but contributions by either a SN with $R > 27.4$ or a
host with $R > 27.6$ (3-$\sigma$ upper limits) cannot be ruled out. If
located at $z = 1.62$, the SN could have been nearly as bright as
expected from appropriately scaling SN1998bw, while a starburst galaxy
as the host would have \mr$ \ga -17$. Proving the presence or absence of
a host galaxy at the OT position or identifying fainter nearby
candidates requires additional deep exposures. If ejection of the
progenitor of GRB\,990510 is considered a possibility, the blue
extended object 12'' east with de-reddened $R = 23.6$ could be a host
candidate.

\acknowledgements{} We thank the ESO staff for the competent
performance of part of the observations in TOO service mode and our
referee, Ralph Wijers, for his helpful comments. KB thanks Dieter
Hartmann for enlightening discussions and comments on GRBs and Wolfram
Kollatschny for helpful comments on starburst galaxies.

\end{document}